\documentclass[12pt]{article}

\usepackage{latexsym,amsmath,amssymb,amsbsy,graphicx}
\usepackage[T2A]{fontenc}
\usepackage[space]{cite} 
\usepackage{epstopdf} 

\allowdisplaybreaks
\begin{document}

\noindent\begin{minipage}{\textwidth}
\begin{center}

{\Large{On a possible method of Casimir pressure renormalization\\ in a ball}}\\[9pt]

{\large A.\,I. Dubikovsky$^{1a}$, P.\,K. Silaev$^{2b}$, O.\,D. Timofeevskaya$^{2}$}\\[6pt]

\parbox{.96\textwidth}{\centering\small\it
$^1$M.\,V.\,Lomonosov Moscow State University, N.\,N.\,Bogoliubov Institute for Theoretical Problems of Microphysics.
Russia, 119991, Moscow, Leninskie Gory, 1.\\
$^2$M.\,V.\,Lomonosov Moscow State University, Physics Department, Chair of Quantum Theory and High Energy Physics. 
Russia, 119991, Moscow, Leninskie Gory, 1., b~2. \\
\ E-mail: $^a$dubikovs@bog.msu.ru,
$^b$silaev@bog.msu.ru}\\[1cc] 

\end{center}

{\parindent5mm 
We propose a method of Casimir pressure renormalization for massive scalar field in a ball.
This method is slightly different from the generally accepted.
An alternative way of choosing the normalization point leads to an exponential pressure dependence on the mass of the field
instead of an inverse polynomial dependence.
The method proposed does not use the scalar quantized field in the exterior domain.
This allows us to bypass the difficulties that appear when one uses the standard approach. 
\vspace{2pt}\par}

\vspace{1pt}\par

\vspace{1pt}\par
\end{minipage}

\section*{Introduction}
\mbox{}\vspace{-\baselineskip}

Back in 1948 H.B.G.Casimir has shown \cite{casimir} that between two plane parallel metallic plates acts the force of attraction,
associated with the presence of zero-point oscillation of the quantized electromagnetic field.
Later the existence of this force has been confirmed experimentally \cite{exp1,exp2}.
Since then, the effect has been studied in detail both theoretically and experimentally \cite{ref1,ref2}.
Besides the direct manifestations and even the possible applications of the effect in nanomechanics
(connected to the change of its sign \cite{repuls}),
it can manifest itself in gravitation theory
\cite{grav1,grav2}, cosmological models \cite{cosmo1,cosmo2}, in the bag model in hadron physics \cite{bag1,bag2} and other applications.
Currently there has been developed quite a number of different methods of exact and approximate calculation of Casimir energy,
 force and pressure \cite{heat,calc1,calc2,calc3}, 
which in the case of simple geometry allow to obtain a finite answer \cite{abel,simple1,simple2,simple3}.
The main difficulty lies in the necessity to perform the renormalization procedure in the original formal expressions for the energy,
force or pressure, which contain divergencies.
For the Casimir force acting between two separate bodies, the renormalization problem is solved trivially, namely
it is enough to use as normalization point the force corresponding to the infinite distance between the bodies, which, obviously,
must be zero.
For the Casimir effect in the case of an isolated body the renormalization problem is much more complicated.
Only for bodies with flat boundaries different methods of renormalization give the same answer.
In the case of curved boundaries the answers in general may be dependent on the method of renormalization \cite{bag1,bag2}.
Even the renormalization for the simplest system, namely quantized field enclosed in a spherical cavity, causes some difficulties.
At present, the generally accepted renormalization method is the following \cite{shar1,shar2,shar3,shar4}:
to a divergent expression corresponding to the Casimir energy (or pressure) inside the sphere
one adds a similar expression, corresponding to the field, quantized outside of the sphere.
In the spaces of odd dimension divergences cancel out, and we obtain a finite answer.
Unfortunately, this procedure is well-founded physically only in the case when we are dealing with a spherical shell of zero thickness.
In this case the answer thus obtained causes no doubts.
In all other cases, one may ask if we might have added to the final finite answer some finite contribution from the field, quantized in the external domain.
Standard examples of situations in which the generally accepted method of renormalization is facing serious difficulties are
the spherical shell of finite thickness, the spherical cavity in a cube, the dielectric sphere placed in a dielectric medium
(in the case where the electric and magnetic medium permeability is arbitrary, not specially selected), etc.   
In this paper we have attempted to construct an alternative renormalization procedure not connected with the external task.
The relevant choice of normalization point will be explained in the next section.

\section{Normalization point choice for Casimir pressure in a ball}
\mbox{}\vspace{-\baselineskip}

Let us first consider the simplest example, namely
a scalar field of mass $m$, defined on the one-dimensional interval $[0,L]$ with zero boundary conditions at the ends of the interval.
We will use the natural system of units $\hbar=c=1$, thus the Lagrangian of the system takes the form
$$ L=\int_0^L dx\; {1\over 2}\left[(\partial_0\varphi)^2-(\partial_1\varphi)^2 -m^2\varphi^2 \right].$$ 
For this model, there exists a large number of methods of regularization and renormalization of pressure or energy
(using $\zeta$-function \cite{zeta}, Abel--Plana formula \cite{abel}, Euclidean Green's function \cite{fgg1,fgg2},
renormalization by differentiation with respect to parameter, etc.), and they all lead to the same result \cite{odnom}.
We will use Euclidean Green function.
Casimir pressure at the right end of the interval $L$
is expressed in terms of the derivative of surface Euclidean Green function $S_\kappa(x,L)$,
which satisfies the equation  
$$ \partial_x^2 S_\kappa (x,L) - \kappa^2 S_\kappa (x,L)=0, $$
and boundary conditions
$$ S_\kappa (0,L)=0,\qquad S_\kappa (L,L)=1.$$
Having $S_\kappa (x,L)$,
we obtain a formal unrenormalized expression for pressure at the right boundary of the interval $L$:
\label{davl}
\begin{equation}
  p(L)=-{1\over 2\pi} \int_0^\infty dt \; {d\over dx} S_{\sqrt{\mathstrut t^2+m^2}}(x,L) \,{\vrule height 3 ex depth 2 ex}_{\; x=L} .\label{davl}
 \end{equation}
The regularization of the resulting expression can be performed, for example, by point splitting,
i.e. we assume $x=L-\epsilon$ instead of $x=L$.
The removal of regularization is achieved by $\epsilon\to 0$.  
Since $${d\over dx} S_\kappa (x,L)=\kappa \, {\rm ch}(\kappa x)/{\rm sh}(\kappa L),$$
the standard renormalization of the integrand consists in
subtracting $\kappa$ from ${d\over dx}S_\kappa(x,L)$: 
\begin{equation}
   {d\over dx}{S\mathstrut}_{\sqrt{\mathstrut t^2+m^2}}^{\rm (ren)}(x,L)={d\over dx}S_{\sqrt{\mathstrut t^2+m^2}}(x,L)-{\sqrt{\mathstrut t^2+m^2}}, \label{reno}
 \end{equation}  
which gives the well-known answer \cite{odnom}: 
$$p(L)=-{1\over \pi}\int_0^\infty dt\; {\sqrt{\mathstrut t^2+m^2}\over \exp(2L\sqrt{\mathstrut t^2+m^2})-1}.$$
This subtraction can be rather well justified.
Indeed,
firstly, the subtracted expression does not depend on the geometry of the system (the length  $L$),
secondly, this subtraction is equivalent to the use of pressure at the border of semi-infinite interval $(-\infty,L]$ as normalization point.
Since the energy of such a semi-infinite system can not depend on the position of the right border of a semi-infinite interval $L$,
the pressure must be equal to zero.
In terms of energy such a subtraction means that we subtract the part of the energy field,
defined on the endless axis, which accounts for the interval $[0,L]$.  
Let us draw attention to the following circumstance.
We have subtracted the local (``surface'') divergence associated with the point $x=L$.
Indeed, the derivative of the surface Green's function corresponding to the interval $(-\infty,L]$
has the form ${d\over dx}{S}^{(0)}_\kappa(x,L)=\kappa\exp(\kappa(x-L))$, so ${d\over dx}{S_\kappa}^{(0)}(L,L)=\kappa$.
The effects associated with the fact the volume is closed are determined by the presence of other points, limiting the domain 
(in the given case, the left boundary of the interval $x=0$), and therefore they are proportional to $\exp(-\kappa a)$,
where $a$ is the typical area size (in the given case, $a=2L$).
Taking into account (\ref{davl}), this leads to an exponential decrease of pressure as the mass of the field increases: $p(L) \sim \exp(-2mL)$.

We now consider the case of a scalar field $\varphi(\vec r)$ with mass $m$, enclosed inside a ball of radius $R$ with zero boundary condition
$\varphi(r=R)=0$. 
Unrenormalized Casimir pressure at the point $\vec x$ on the boundary may be written in exactly the same way,
as in the one-dimensional case:
\begin{equation}
  p(\vec x)=-{1\over 2\pi}\int_0^\infty \; dt \;  \left(\vec {n\vphantom{\nabla}} \cdot \vec{\nabla}_{\vec y}\right) \; S_{\sqrt{\mathstrut t^2+m^2}} (\vec y,\vec x)\,{\vrule height 3ex depth 2 ex}_{\;\vec y=\vec x}\, ,
  \label{davl15}
  \end{equation} 
where $\vec n$ is normal to boundary surface at the point $\vec x$, and $S_\kappa(\vec y ,\vec x)$ is Euclidean surface Green's function,
satisfying the equation:
$$ \triangle_{\vec y} S_\kappa(\vec y ,\vec x) -\kappa^2 S_\kappa(\vec y ,\vec x)=0. $$  
We use the expression for surface Green's function in a ball in the form of an expansion in spherical harmonics:
$$
S_{\kappa} (\vec y,\vec x)=\sum_{lm}Y_{lm}(\theta_y,\varphi_y)
Y_{lm}^*(\theta_x,\varphi_x) 
\frac{\sqrt{R} I_{l+1/2}(\kappa r_y)}{R^2\sqrt{r_y}  I_{l+1/2}(\kappa R)},
$$
where $r_x=R$, and $I_{l+1/2}(z)$ is Infeld function. 
Later we will need expressions for the asymptotics of Infeld and Macdonald functions $I_\nu,K_\nu$ for large $\nu$.
They have the following form \cite{batem}:
$$ I_\nu(z)\sim (z^2+\nu^2)^{-1/4}\exp(\sqrt{z^2+\nu^2}-\nu \;\hbox{arsh}(\nu/z)),  $$
$$ K_\nu(z)\sim (z^2+\nu^2)^{-1/4}\exp(-\sqrt{z^2+\nu^2}+\nu \;\hbox{arsh}(\nu/z)),  $$
i.e. Infeld function $I_\nu$ decreases and the Macdonald function $K_\nu$ increases with increasing~$\nu$.
It is convenient to apply here spherical Infeld and Macdonald functions
$i_l(x)\equiv I_{l+1/2}(x)\sqrt{\pi/(2x)}$, $ k_l(x)\equiv K_{l+1/2}(x)\sqrt{2/(\pi x)}$.
Substituting the expression for $S$ into (\ref{davl15}), we get the following formal answer:
\begin{equation}
  p(R)=-{1\over 2\pi} \int_0^\infty \; dt \; \sum_{l=0}^\infty \frac{2l+1}{4\pi}\; {\sqrt{\mathstrut t^2+m^2}\; {i\,}'_{l}(\sqrt{\mathstrut t^2+m^2}R) \over R^2\; i_{l}(\sqrt{\mathstrut t^2+m^2}R) } .  \label{davl2}
 \end{equation} 
Unlike the one-dimensional case, where the integrand was finite,
here the derivative of surface Green's function is singular, since the sum over $l$ diverges.
Let us regularize the integrand as follows.
We multiply each term in the sum by $f(l+1/2)$, where  
$$f(m)=(1+\epsilon m)^{-2m-n}.$$ 
Here $n$ is integer and $n>3$. 
Removal of regularization is achieved by $\epsilon\to 0$. 
In fact, to ensure convergence of the sum over $l$ such a regularization is excessive, as the terms have the asymptotic $O(l^2)$, 
yet it will further prove necessary to perform the renormalization procedure. 

We now turn to the problem of renormalization of the regularized expression obtained. 
We first note that, unlike the one-dimensional case, the divergent part of the expression depends on the geometry of the system. 
This is due to the fact that the surface divergence is determined by boundary curvature, i.e. depends on the radius $R$.
Indeed, let us consider the one-dimensional problem corresponding to given spherical harmonics $l$.
Unlike the one-dimensional case, in this problem the appearance of the centrifugal potential breaks the translational invariance of the problem.
Therefore the renormalization term can depend on the position of the boundary, i.e., the radius $R$. 

The standard renormalization procedure consists in adding terms corresponding to the exterior problem,
i.e. the scalar field quantized outside the sphere of radius $R$.
For the exterior problem the divergent term associated with the curvature of the border is of opposite sign,
so we obtain a finite expression for the pressure.
However, as was mentioned above, this procedure has several disadvantages.
Besides the already mentioned difficulties,
we note the fact that it gives an answer that decreases with increasing mass of the field as inverse polynomial.
It seems that such a nature of decrease is not quite reasonable from purely physical considerations.

Indeed, let us consider a cube with edge $L$ with a scalar field of mass $m$ inside it.
It is well known that Casimir energy as well as pressure in this case decreases exponentially with increasing mass of the field.
Moreover, for the cube (as well as for any rectangular parallelepiped) the renormalization result is beyond doubt.
A wide variety of methods of regularization and renormalization \cite{ref1}, \cite{ref2} lead to the same answer.
Therefore, in the further considerations we will use the answer for the cube, and all the answers for domains with flat boundaries,
as a kind of reference point.

The divergence of surface Green's function for any point on a face of the cube that is far from edges of the cube
(i.e. from the other faces of the cube)
coincides exactly with the divergence of surface Green's function for the plane, i.e., does not depend on $L$,
and the influence of other faces decreases exponentially with increasing mass of the field.
Now let us deform the cube so that in the neighborhood of the point considered the curvature of the surface became low,
but non-zero (let, for example, both radii of curvature correspond to a sphere with large but finite radius $R$).
This small deformation leads to a dramatic change in the nature of Casimir pressure decrease as the mass of the field $m$ increases.
The exponent $\exp(-2mL)$ will be replaced by the inverse polynomial dependence.
From a mathematical point of view it is quite acceptable, but it seems that this outcome is not acceptable from a purely physical point of view.

Exactly the same arguments can be repeated for two planes separated by a distance $L$.
Again, the answer for the case of zero curvature causes no doubt, and turns out to be proportional to $\exp(-2mL)$. 
Can a small deformation, leading to a small but nonzero curvature of the surface,
lead to a substantial change in the nature of dependence on the mass of the field?
We emphasize once again that from a mathematical point of view it certainly can. 
But the renormalization procedure requires a particular choice of normalization point,
i.e. of certain conditions which must be satisfied by the renormalized answer.
It seems a physically natural requirement
that a small deformation of the surface did not result into a radical change in the renormalized answer obtained.
As a reference point we choose the answers corresponding to flat boundaries,
as for flat boundaries, firstly, the procedure for removing surface divergence is beyond doubt,
and, secondly, the answers obtained do not lead to any further difficulties.
Neither rectangular pistons in a rectangular parallelepiped,
nor rectangular parallelepiped with walls of finite thickness does not cause those difficulties,
that arise for a sphere with finite wall thickness.
Our demand for a kind of ``continuity'' is certainly arbitrary, but it seems physically justified.

Thus we will try to build a renormalization procedure,
for which a small deformation of the cube surface would not lead to radical changes in the answer for Casimir pressure.
In other words, we assume that the terms which do not decrease exponentially as the mass of the field increases,
should be attributed to local terms related to the point in question, and include them into the surface divergence, 
while all the terms that decay exponentially,
will be viewed as a result of influence of other (remote) boundary points on the point under consideration and
that they are regarded as the renormalized answer for Casimir pressure.

We introduce the notation $y=R\sqrt{\mathstrut t^2+m^2}$
and split the regularized sum into two sums $S_1$ and $S_2$,
which are the sums in even and in odd $l$:
$$
S_1=\sum_{l=2k} s_1(l), \qquad S_2=\sum_{l=2k+1} s_2(l), $$
\begin{equation}
s_1(l)=s_2(l)={2l+1 \over 4\pi}\; {y\over R^3}\; { i'_{l}(y) \over   i_{l}(y) }\; f(l+1/2) .  
\label{kuski}
\end{equation}
The meaning of such a splitting will become apparent later. 

Now we take into account that 
$$  I'_\nu (x)=(I_{\nu +1}(x)+I_{\nu -1}(x))/2,  $$
and the exponentially decreasing term in $I_{l+1/2}(x)$ is equal to
$$   (-1)^{l+1}K_{l+1/2}(x)/\pi . $$
We keep only the terms proportional to $\exp(-2y)$ in each  $s_1(l)$.  We get
$$  s_1^{\rm (ren)}(l)= {2l+1 \over 4\pi}\; {y\over R^3}\; {( k_{l+1}(y)+ k_{l-1}(y)) \over  2 \; i_{l}(y) }\; f(l+1/2)=  $$
$$  = {2l+1 \over 4\pi}\; {y\over R^3}\; { K_{l+3/2}(y)+ K_{l-1/2}(y) \over \pi I_{l+1/2}(y) }\; f(l+1/2).$$
Such a renormalization is completely analogous to the one-dimensional renormalization
$$ \frac{\kappa\; {\rm ch} (\kappa L)}{{\rm sh} (\kappa L)} \to \frac{\kappa\; \big( \exp(-\kappa L)+\exp(-\kappa L)\;\big)}{2\;{\rm sh} (\kappa L)} =\frac{2\kappa \exp (-\kappa L)}{\exp (\kappa L)-\exp (-\kappa L) },   $$
accomplished in (\ref{reno}).

In an absolutely similar way, each term in the sum on odd $l$ is replaced by
$$ s_2^{\rm (ren)}(l)= {2l+1 \over 4\pi}\; {y\over R^3}\; { -K_{l+3/2}(y)- K_{l-1/2}(y) \over \pi I_{l+1/2}(y) }\; f(l+1/2).$$
Note that the sign in this expression differs from the sign for even $l$.

Now consider procedure of regularization removal.
Here we encounter purely technical difficulties related to the fact that renormalized sums converge worse than the original sum.
The summands of the renormalized sum have asymptotics $[l!(2/y)^l]^2$.
It is for this reason that we used excess regularization for our original sum.
In direct calculation of the renormalized sums $S_1^{\rm ren}$ and $S_2^{\rm ren}$
convergence is achieved only when the values of regularization parameter are $\epsilon>2/(ye)$,
so the direct removal of regularization by  $\epsilon\to 0$ is impossible.

This difficulty can be easily resolved by using analytic continuation of the renormalized sum in the parameter of regularization $\epsilon$. 
Indeed, for $\epsilon>2/(ye)$ for each of the renormalized sums (\ref{kuski}) one could write the following integral representation:
$$ S_1^{\rm (ren)}= \sum_{k=0}^\infty {4k+1 \over 4\pi} {y\over R^3} { K_{2k+3/2}(y)+ K_{2k-1/2}(y) \over \pi I_{2k+1/2}(y) } f(2k+1/2) = $$ 
$$
={i\over 2}
\int_{-\infty}^\infty  \;dx\; {\tanh(\pi x)}\;{4ix-1 \over 4\pi}\;{y\over R^3}\; {K_{2ix+1/2}(y)+K_{2ix-3/2}(y) 
\over \pi  I_{2ix-1/2}(y) }\; f(2ix-1/2), $$
for even $l$, and likewise for odd $l$:
$$ S_2^{\rm (ren)}= \sum_{k=0}^\infty  {y\over R^3}\; {4k+3 \over 4\pi}\; { -K_{2k+5/2}(y)- K_{2k+1/2}(y) \over \pi I_{2k+3/2}(y) }\; f(2k+3/2) = $$ 
$$
=-{i\over 2}
\int_{-\infty}^\infty  \;dx\; {\tanh(\pi x)}\;{4ix+1 \over 4\pi}\;{y\over R^3}\; {K_{2ix+3/2}(y)+K_{2ix-1/2}(y) 
\over  \pi I_{2ix+1/2}(y) }\; f(2ix+1/2). $$
Both representations are obtained by closing the contour of integration by an infinite semicircle to the right of the imaginary axis.
In the integral representation there is a smooth limit at $\epsilon\to 0$, so the regularization can be removed.

The final expression for the renormalized pressure (\ref{davl2}) will take the form 
$$ p(R)=-{1\over 2\pi} \int_0^\infty \; dt \;  
 {i\over 2}\;{y(t)\over R^3} \times  \hbox to 8 cm{\hfill} $$ $$
\times \int_{-\infty}^\infty  \;dx\; {\tanh(\pi x)} \left[{4ix-1 \over 4\pi}\; {K_{2ix-3/2}(y(t))+K_{2ix+1/2}(y(t)) 
\over  \pi I_{2ix-1/2}(y(t)) } \; - \right. $$
$$ \hbox to 3 cm{\hfill} \left.  - \; {4ix+1 \over 4\pi}\; {K_{2ix-1/2}(y(t))+K_{2ix+3/2}(y(t))
\over  \pi I_{2ix+1/2}(y(t)) }  \right] , $$
where $y(t)=R\sqrt{\mathstrut t^2+m^2}$.

This expression proves to be finite and decreases exponentially with increasing mass $m$ of the field.  
      
\begin{figure}[tbh!]
\centerline{\includegraphics{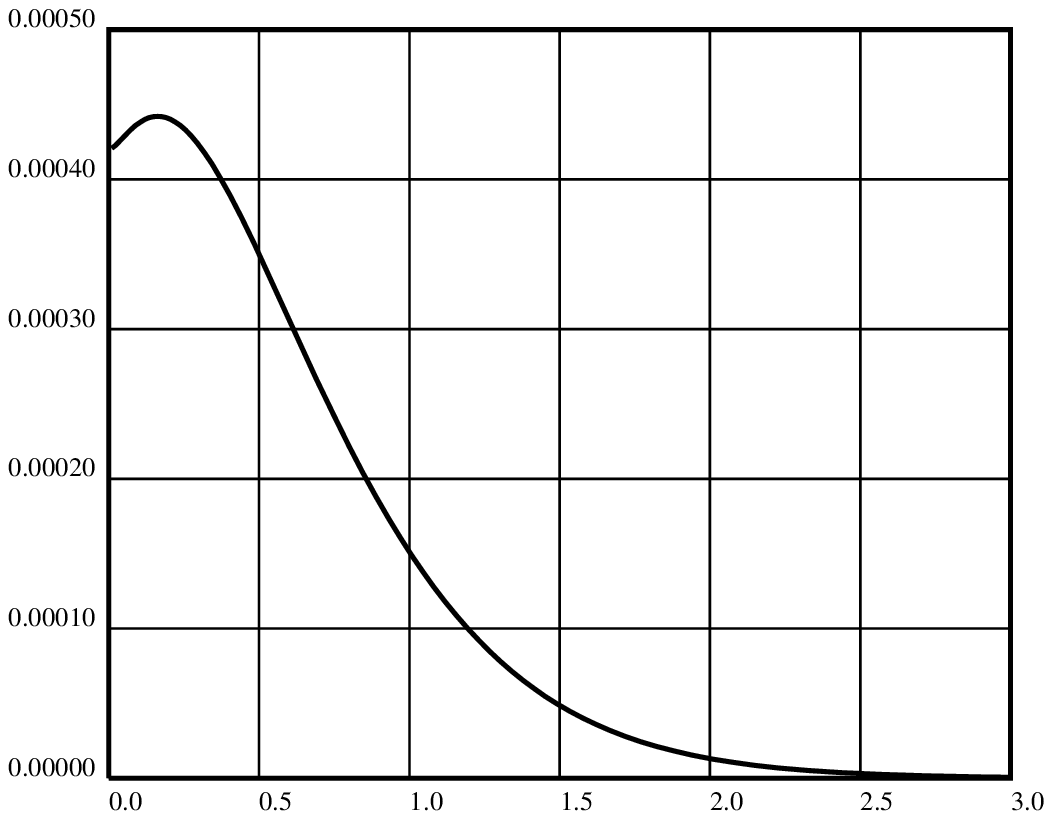}}
\caption{The dependence of the pressure $p(R)$ on the mass $m$ for the field in a sphere of unit radius}
\label{fig1}
\end{figure}

The pressure dependence of the mass of the field for a unit radius sphere $R=1$ is shown on Fig.~\ref{fig1}.
To give here the graph of pressure renormalized in the standard way is to no purpose,
since the nature of decrease is different (inverse polynomial instead of exponential).
However, the comparison of the answers for a massless field is of certain interest:
by the standard method we obtain $0.00282$, while our answer is $0.000421$ (contribution of even $l$ is $0.003175$, whereas contribution of odd $l$ is $-0.002754$).
The average pressure on the face of the cube with unit edges is equal to $-0.00524$,
and the average pressure on the face of a cube with unit ``radius'' (i.e., with the edge 2) is equal to $-0.0013$.

\section*{Conclusions}
\mbox{}\vspace{-\baselineskip}

The method of renormalization proposed, strictly speaking, includes similar arbitrary actions as the standard method.
While subtracting the infinite contributions there inevitably arises an arbitrariness,
which can be removed only with the involvement of physical considerations about a particular choice of the normalization point.
It seems that the inclusion in the surface divergence of all the terms that do not decrease exponentially with increasing mass of the field
is more physical than adding to a divergent answer the answer to a different (external) problem. 
Since one of the methods of describing Casimir effect is a picture of a virtual particle that is born,
reflected from the domain's wall and then absorbed, the dependence of the effect on the mass of the field of form $\exp(-ma)$,
where $a$ is the typical size of the region seems the most natural.
Furthermore, the method we proposed is free from certain disadvantages of the standard method.
Indeed, we have renormalized exactly the original (internal) problem for the scalar field,
with absolutely no consideration of the external problem.
This is why the method can be applied in the case of a spherical shell of finite thickness, as well as for a spherical cavity truncated in a cube,
as well as for dielectric sphere with an arbitrary dielectric permittivity.

On the other hand, the proposed method requires exclusively ``strong'' (excessive) regularization,
so the class of admissible regularizing functions is noticeably narrower than for the standard method.
Moreover, the proposed method requires subsequent analytic continuation of obtained renormalized expression.
Finally, it essentially uses the spherical symmetry of the system in question --- the renormalization of pressure for arbitrary surface, i.e. for the case of two different radii of curvature  requires  additional consideration.

In conclusion, the authors consider it a pleasant duty to thank O.\,V.~Pavlovsky, M.\,V.~Ulybyshev and K.\,A.~Sveshnikov
for attention to this work and fruitful discussions.

\newpage

\end{document}